# Summary of the DISPLACE Challenge 2023 - DIarization of SPeaker and LAnguage in Conversational Environments

Shikha Baghel[1]*, Shreyas Ramoji[1], Somil Jain[2], Pratik Roy Chowdhuri[2], Prachi Singh[1], Deepu Vijayasenan[2], Sriram Ganapathy[1]

[1]*LEAP Lab, Department of Electrical Engineering, Indian Institute of Science, Bengaluru, India*
[2]*Department of Electronics and Communication, National Institute of Technology Karnataka, Surathkal, India*

**Abstract**

In multi-lingual societies, where multiple languages are spoken in a small geographic vicinity, informal conversations often involve mix of languages. Existing speech technologies may be inefficient in extracting information from such conversations, where the speech data is rich in diversity with multiple languages and speakers. The **DISPLACE** (DIarization of SPeaker and LAnguage in Conversational Environments) challenge constitutes an open-call for evaluating and bench-marking the speaker and language diarization technologies on this challenging condition. The challenge entailed two tracks: Track-1 focused on speaker diarization (SD) in multilingual situations while, Track-2 addressed the language diarization (LD) in a multi-speaker scenario. Both the tracks were evaluated using the same underlying audio data. To facilitate this evaluation, a real-world dataset featuring multilingual, multi-speaker conversational far-field speech was recorded and distributed. Furthermore, a baseline system was made available for both SD and LD task which mimicked the state-of-art in these tasks. The challenge garnered a total of 42 world-wide registrations and received a total of 19 combined submissions for Track-1 and Track-2. This paper describes the challenge, details of the datasets, tasks, and the baseline system. Additionally, the paper provides a concise overview of the submitted systems in both tracks, with an emphasis given to the top performing systems. The paper also presents insights and future perspectives for SD and LD tasks, focusing on the key challenges that the systems need to overcome before wide-spread commercial deployment on such conversations.

## 1. Introduction

Multilingualism, the expression of multiple languages, is increasingly common in many parts of the world [1, 2]. In this context, *code-mixing* represents the intra-sentential switching [3], where words or short phrases from one language (secondary) are used within an utterance[2] of another language [4]. Similarly, *code-switching* involves inter-sentential switching [3], where language switching occurs at the utterance level[3].

---

*Corresponding author
  *Email addresses:* `shikhabaghel@iisc.ac.in` (Shikha Baghel[1]), `shreyasr@iisc.ac.in` (Shreyas Ramoji[1]), `jrf.somiljain@nitk.edu.in` (Somil Jain[2]), `pratikranjanroychowdhuri.217ec009@nitk.edu.in` (Pratik Roy Chowdhuri[2]), `prachisingh@iisc.ac.in` (Prachi Singh[1]), `deepuv@nitk.edu.in` (Deepu Vijayasenan[2]), `sriramg@iisc.ac.in` (Sriram Ganapathy[1])

  [2]For example, कल मैं एक **movie** देखने जा रहा हूँ। (*Tomorrow I am going to watch a movie*). Here, the English word "movie" is used in a Hindi sentence.
  [3]For example, आज मैं बहुत खुश हूँ। **Let's go for a party.** (*Today I am very happy. Let's go for a party*)

Further, multi-speaker audio in conversational settings elicits unstructured and challenging conditions for analytics and understanding.

The presence of code-mixed or code-switched speech is a very common phenomenon in bilingual or multilingual communities [5, 6], across several locales [7–9]. In North America, the use of Spanish with English has grown significantly [10, 11] while the code-switching of French-German in Switzerland [12] and Frisian-Dutch in the Netherlands [13] are prominent examples of bilingual societies in Europe. In North Africa, the dominant use of Arabic-English in Egypt [14] and French-Arabic in Algeria, Morocco, and Tunisia [15] are other examples of code-switching societies. South Africa has bilingual communities, namely Sepedi-English, English-isiZulu, English-isiXhosa, English-Setswana, and English-Sesotho [16]. In Hong Kong, the embedding of English words in colloquial Cantonese [17] is common, while mix of Mandarin and Taiwanese is a common phenomenon in Taiwan [18]. Similarly, a combination of English and Mandarin is used in Singapore and Malaysia [6]. India is one of the most prominent examples of a multilingual society with 23 official languages, including Indian English. There is frequent usage of English words or phrases while conversing in other native languages [2].

In multilingual societies, developing speech technology for a broad spectrum of population is challenging for applications like content analysis, summarization [19], automatic speech recognition (ASR) [2, 20], speech synthesis [21] and language diarization [22]. Most state-of-the-art (SOTA) speech systems, such as ASR and diarization, perform very well on monolingual speech. However, such systems are inefficient in handling multi-speaker code-switched/mixed audio [23]. Therefore, automatic extraction of speaker and language information is an essential pre-processing task for any speech-based solution [23].

This paper focuses on the speaker and language diarization in conversational settings containing frequent code-mixing and code-switching. The *Diarization* term is extensively associated with speaker segmentation in the literature, where the goal is to segment the audio signal into speaker-homogeneous segments. In recent years, researchers have also been interested in diarizing audio signals based on the spoken language, which is defined as language diarization (LD). The LD task provides the answer of "which language was spoken when". Existing SD systems are mainly developed for monolingual settings. Simultaneously, language recognition systems are mainly developed for situations where a speaker speaks only in one language. This paper presents a summary of the DISPLACE challenge[4], which was organized with the goal of benchmarking and streamlining efforts for improving diarization solutions in multi-lingual social conversations. The paper details three aspects - i) Data collection and annotation, ii) Baseline system development using state-of-art techniques and, iii) Key contributions from the challenge participants which improved the performance on the SD/LD tasks.

## 2. Related Works

***Speaker Diarization*** - The National Institute of Standards and Technology-Rich Transcription (NIST-RT) [33, 34] evaluations played a crucial role in advancing the field of SD by providing evaluation benchmarks, introducing the Diarization Error Rate (DER) metric, and fostering diarization and speaker recognition research. Multiple NIST-RT evaluations were organized from 2002 to 2009, with SD as one of the metadata extraction tasks. In NIST evaluations, SD systems were mainly developed and evaluated on Broadcast News (BN), conversational telephonic speech, conference and lecture room meetings conducted in English [33–36]. Other popular datasets used for SD task are ICSI [24], AMI [25], CHiME-5/6 [26, 27], LibriCSS [28], IberSpeech-RTVE 2018 [29], VoxConverse [30], DIHARD-III [31] and CALLHOME [32]. A summary of the datasets used for the SD task is provided in Table 1.

ICSI and AMI datasets are mainly focused on formal meeting scenarios. The issues related to distant multi-microphone diarization in daily life conversations are addressed in CHiME-5/6 challenges [26, 27]. SD task in TV shows is addressed in IberSpeech-RTVE challenge [29], a part of the Albayzin evaluation series. Recent challenges, such as VoxCeleb Speaker Recognition Challenge 2020 (VoxSRC-20) [37] and DIHARD [31] foster the SD research in more complex and unconstrained scenarios. The DIHARD challenges [31, 38, 39] were organized with the goal of developing SD systems which are robust to different conversational domains, demographics, recording ambience, and set up, by addressing challenging scenarios, such as child recordings, clinical interviews, reverberant speech, audio from wearable devices in outdoor settings, etc.

---

[4]https://displace2023.github.io

Table 1: A summary of speech corpora used for the speaker diarization (SD) task. Information of the listed datasets is given in terms of speech style, multilingualism, presence of code-switched instances, number of speakers per recording (# Spkrs/recording), dataset size in hours, presence of language (Lang.) turn and content annotation. Here, "NA" stands for Not Applicable.

| Dataset | Style | Multilingual | Code-switch | # Spkrs / recording | Size (hrs) | Lang. turn annotation | Content annotation |
|---|---|---|---|---|---|---|---|
| ICSI [24] | Formal Meeting | ✗ | ✗ | 3-10 | 72 | NA | ✓ |
| AMI [25] | Formal Meeting | ✗ | ✗ | 3-5 | 100 | NA | ✓ |
| CHiME-5/6 [26, 27] | Informal conversation | ✗ | ✗ | 4 | 50 | NA | ✓ |
| LibriCSS [28] | Simulated conversation | ✗ | ✗ | 8 | 10 | NA | ✓ |
| IberSpeech-RTVE [29] | TV shows | ✗ | ✗ | - | 38 | NA | ✓ |
| VoxConverse [30] | Mixed of styles | ✗ | ✗ | 1–21 | 74 | NA | ✗ |
| DIHARD-III [31] | Mixed of styles | ✗ | ✗ | 1–10 | 67 | ✗ | ✗ |
| CALLHOME [32] | Informal conversation | ✓ | ✗ | 2-7 | 20 | ✗ | ✓ |
| DISPLACE [23] | Informal conversation | ✓ | ✓ | 3-5 | 32 | ✓ | ✓ |

All these datasets are monolingual. A multilingual dataset, CALLHOME, is also available for the SD task. The CALLHOME corpus contains six different languages but only one language per recording. In contrast, the DISPLACE corpus stands out for its unique composition of multilingual recordings featuring frequent instances of code-mixing and code-switching.

**Language Diarization** - In multilingual speech processing, LD has emerged as an essential task [46, 47]. Due to the lack of publicly available datasets, researchers developed new corpora for LD. Some of the popularly used code-mixed/switched datasets are summarized in Table 2. One of the popular LD datasets is South-East-Asia Mandarin/English (SEAME) [6, 42]. This dataset has been used in many works [9, 22, 47]. Other datasets are FAME (Frisian Audio Mining Enterprise) [43], Chinese-English COde-switching Speech (CECOS) [41], OC16-CE80 [20], South African Sap Opera (SASO) [16]. In recent years, speech datasets containing code-switched instances for Indian languages have also been developed. Some of these datasets are MSR Hindi-English [44], IITG-HingCoS [2] and WSTCSMC [45].

All the current datasets consist of two languages per recording, with a total of two languages in each dataset, except for SASO and WSTCSMC. In contrast, the DISPLACE corpus includes recordings with monolingual, bilingual, and trilingual content, encompassing seven different languages. Moreover, the existing datasets feature a maximum of two speakers per recording. However, the DISPLACE corpus includes recordings with 3 to 5 speakers. Most code-switched datasets were primarily developed for enhancing ASR systems and therefore include transcriptions. To the best of our knowledge, none of these datasets, except SASO, provide speaker labels, which limits their utility for applications such as speaker diarization and speaker recognition. In contrast, the DISPLACE dataset offers speaker labels in addition to language annotations and transcriptions.

Table 2: An overview of code-switched and code-mixed speech corpora. Details about the listed datasets are provided in terms of speech style, number of languages per recording (# Lang./file), total number of languages present in the dataset (Total # lang.), number of speakers per recording (# Spkrs/recording), total number of speakers present in the dataset (Total # spkrs), dataset size in hours, presence of speaker (Spkr) turns and content annotations.

| Dataset | Style | # Lang. /file | Total # lang. | # Spkrs / file | Total # spkrs | Size (hr) | Spkr turn annotation | Content annotation |
|---|---|---|---|---|---|---|---|---|
| CUMIX [40] | Read | 2 | 2 | 1 | 80 | 17.0 | - | ✓ |
| CECOS [41] | Read (utterance) | 2 | 2 | 1 | 77 | 12.1 | - | ✓ |
| SEAME [42] | Conversation & interview | 2 | 2 | 2 | 157 | 63.0 | - | ✓ |
| FAME [43] | Radio broadcast | 2 | 2 | - | 542 | 19.0 | - | ✓ |
| OC16-CE80 [20] | Read (utterance) | 2 | 2 | 1 | 1500 | 80.0 | - | ✓ |
| SASO [16] | TV episodes | 2 | 5 | - | - | 14.3 | ✓ | ✓ |
| MSR Hindi-English [44] | Conversation | 2 | 2 | 2 | 535 | 50.0 | - | ✓ |
| IITG-HingCoS [2] | Read (utterance) | 2 | 2 | 1 | 101 | 25.0 | - | ✓ |
| WSTCSMC [45] | Read & Conversation | 2 | 4 | - | - | 60.0 | - | ✓ |
| DISPLACE [23] | Conversation | 1, 2 & 3 | 7 | 3-5 | 165 | 32.0 | ✓ | ✓ |

## 2.1. Contributions

Compared to the datasets currently available, the DISPLACE corpus introduces several unique characteristics. Firstly, it encompasses conversations in monolingual, bilingual, and trilingual settings with frequent code-mixed and code-switched instances. Secondly, it features recordings in seven distinct languages in social conversational settings. Thirdly, each recording includes 3-5 speakers, a contrast to the existing LD datasets where there are a maximum of 2 speakers per recording. Lastly, speaker and language turn annotations are manually generated for evaluating and bench-marking automatic systems. Many of the current language diarization models assume that the test recording contains two known languages [8, 22, 48]. However, the DISPLACE challenge tackles a broader scenario by involving unknown languages during the testing phase.

The present work describes our efforts in organizing the DISPLACE (DIarization of SPeaker and LAnguage in Conversational Environments) challenge at INTERSPEECH-2023 [23]. The main contributions of this paper are,

1. The challenge entails a first-of-its-kind task to perform speaker and language diarization on the same data containing multi-speaker social conversations in multilingual code-mixed speech. In this paper, we describe the data recording protocols used to ensure a naturalistic real-world data setting.

2. The DISPLACE corpus consists of 32 hrs of natural conversations involving multiple speakers, bilingual and tri-lingual code-mixed and code-switched conversations in seven Indian languages. The corpus includes annotations for speaker-turns, language-turns, and transcriptions[5]. This paper provides a detailed analysis and insights into the dataset with key statistics related to speaker and language turns.

---
[5]Initially, only speaker-turn and language-turn annotations were released to the participating teams as part of the DISPLACE challenge, but the DEV set of the corpus, along with its transcriptions, is made publicly accessible at https://zenodo.org/record/8375161

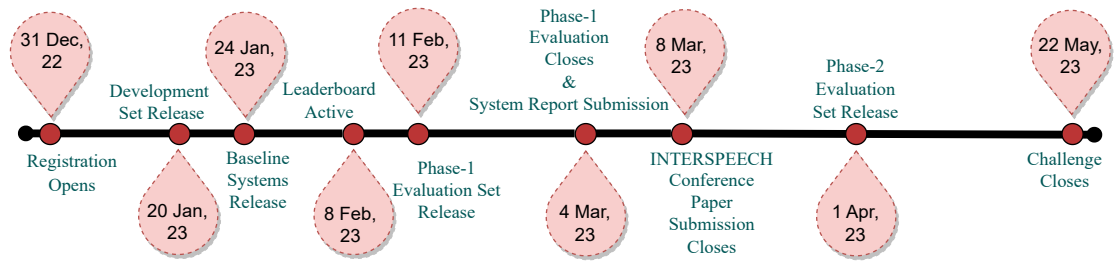

Figure 1: The DISPLACE challenge timeline.

3. This paper offers a comprehensive account of the baseline approaches and the systems that were submitted to the challenge, with a particular focus on the top three systems from both tracks. The paper also sheds light on the sources of errors, thereby paving the way for future research avenues.

## 3. DISPLACE Challenge

The DISPLACE challenge[6] commenced on $31^{st}$ December, 2022 and concluded on $22^{nd}$ May, 2023. The challenge had two tracks: Speaker Diarization (SD) and Language Diarization (LD). A development set, evaluation set, and baseline systems for both tracks were released to all the registered teams. A timeline of the challenge is displayed in Figure 1. Two separate evaluation sets, i.e., Eval-Phase1 and Eval-Phase2, were released. System submissions for both the development and evaluation sets were made using a leaderboard[7]. A total of 42 teams from academia and industry registered for the challenge. Among these, 13 and 6 teams submitted their final system to the SD and LD tracks, respectively.

### 3.1. DISPLACE Dataset

The DISPLACE corpus comprises informal conversations of duration 30-60 minutes, involving 3 to 5 speakers. The participants provided meta-information, including age, gender, education, native place, mother tongue, languages known, and proficiency level in English. Those who reported proficiency in at least one Indian language (L1) and Indian English were selected. For a conversation, participants were chosen based on their L1. Topics for the discussions were selected through mutual agreement among the participants and covered various topics of social interest, entertainment, festivals, technology, climate change, and more. Before each recording, participants submitted their consent in written form and received a briefing on the guidelines. Monetary compensation was provided to all participants for their contribution. The DISPLACE corpus contains 32 hours of data collected from two academic institutes: IISc[8] and NITK[9].

#### 3.1.1. Recording Setup

The data is collected in two types of recording setups. The recording rooms used, varied in terms of their shape, size, and acoustic properties. The absence of dedicated soundproofing in these recording rooms resulted in recordings that capture natural noise and background speech. Consequently, recordings contain outside corridor noise, including voices of external individuals, drilling sounds, and telephone rings. Overall, both recording setups were similar with only minor differences.

---

[6]https://displace2023.github.io/
[7]https://codalab.lisn.upsaclay.fr/competitions/10588
[8]Indian Institute of Science, Bengaluru, India
[9]National Institute of Technology Karnataka, Surathkal, India

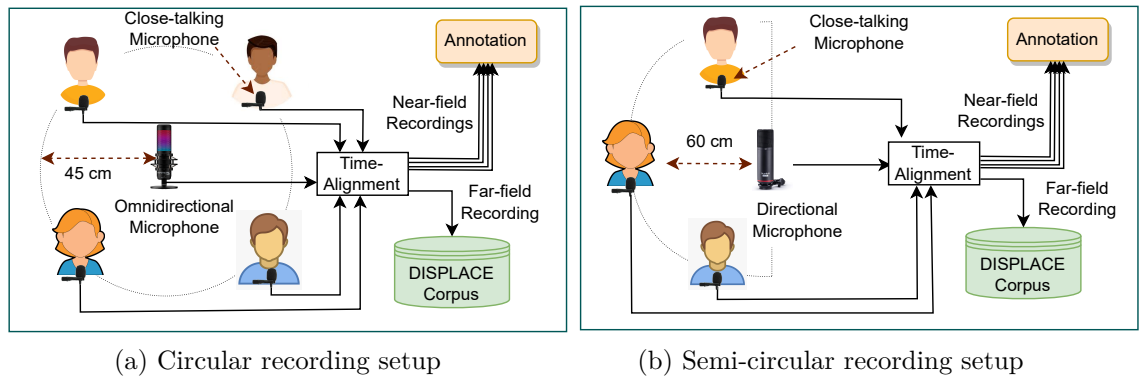

(a) Circular recording setup     (b) Semi-circular recording setup

Figure 2: Schematic of the recording setups employed at (a) IISc and (b) NITK. The circular recording setup (a) involves an omnidirectional far-field microphone, whereas a unidirectional far-field microphone is used in the semi-circular setup (b).

A pictorial illustration of the recording setups used at IISc and NITK are shown in figure 2 (a) and (b), respectively. The data collection paradigm involves a close-talking lapel microphone for each participant and a shared far-field desktop microphone positioned at the center of the table. A lapel microphone, connected to either an audio recorder (figure 2 (a)) or an Android phone (figure 2 (b)), was employed to record the close-talking audio. Uniform settings were maintained for all the audio recorders (or Android phones) throughout all the conversations. The far-field recordings were captured through an omnidirectional (figure 2 (a)) or unidirectional (figure 2 (b)) far-field microphone. For the recordings, speakers were placed in either a circular or semi-circular arrangement, as shown in figure 2 (a) and (b), respectively. In the circular seating setup, speakers were positioned at approximately uniform intervals from one another, maintaining a distance of around 45 cm from the far-field microphone. In the semi-circular seating arrangement, participants were equidistant from each other and maintained a distance of about 60 cm from the unidirectional far-field microphone.

The close-talking speech was captured for annotation purposes, making it easier for human annotators to identify speech, speaker, and language boundaries. However, the system development and evaluation was performed using the far-field audio data.

*3.1.2. Audio Specifications*

The audio (close and far-field) recordings are single-channel wav files sampled at 16 kHz. For each conversation, all the close-talking recordings were time-synchronized and normalized.

*3.1.3. Annotation Protocol*

The data annotations were produced by professional annotators who listened to the close-talking recordings and generated the required annotations for the target speaker (the speaker wearing the lapel microphone). For each conversation, three different kinds of annotations were generated, viz. speech vs. non-speech, language, and content.

- **Speech vs. non-speech annotations:** The initial annotation task involved identifying the speech activity of the target speaker (speaker activity). The other remaining regions, including silence, long-pause, and speech regions of non-target speakers were labeled as non-speech. The prominent non-speech audio instances of the target speaker, such as breathing, laughing, lip-smacking, coughing, etc., were also marked. Evident background sounds, such as telephone rings and any other noise, were marked separately. The annotation codes used for these categories are detailed in Table 3.

- **Language annotations:** The second annotation task focused on identifying language labels corresponding to the target speaker activity regions. A language label was assigned for each word. During informal conversations, it is observed that participants use a lot of fillers and back-channel words, such as "ahh", "umm", "uh-umm", "ooo", "ahaa", etc. Past and future context across the non-lexical words were used for deciding the language label of these filler words. The language codes used for this task are reported in table 4.

Table 3: Annotation codes used for non-speech acoustic categories encountered in the dataset.

| Code | Description |
| --- | --- |
| [*breath*] | inhalation and exhalation between words, yawning |
| [*lipsmack*] | coughing, throat clearing, sneezing |
| [*laugh*] | laughing, chuckling |
| [*ring*] | telephone ring |
| [*other*] | any other prominent sound |

Table 4: Language codes used for code-mixed annotations.

| Language | Code | Language | Code |
| --- | --- | --- | --- |
| English | en | Hindi | hi |
| Telugu | te | Kannada | ka |
| Tamil | ta | Malayalam | ma |
| Bangla | ba | | |

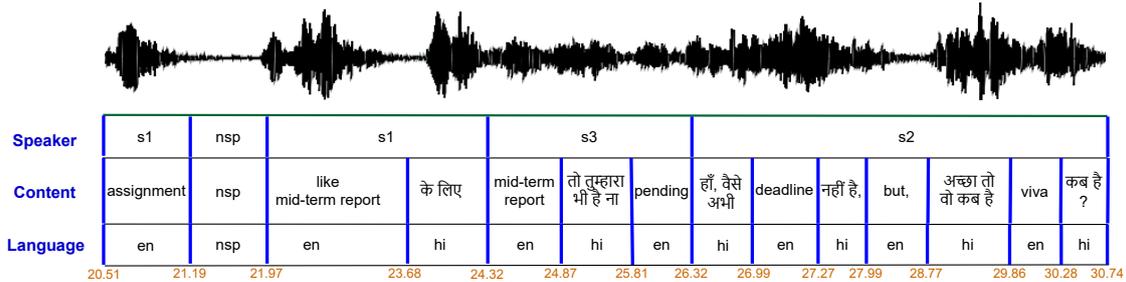

Figure 3: A sample from the DISPLACE corpus showcases code-mixed instances in Hindi-English, with the first tier indicating speaker-turn labels. The content tier presents transcriptions in the native script, while the third tier shows language labels.

- **Content Annotations:** The third annotation task aimed to produce multilingual transcripts of the content spoken by the target speaker. These transcriptions are part of the current release of the dataset.

The annotations underwent multi-level quality checks. The individual lapel microphone annotations were combined to generate a single annotation file for each conversation. Figure 3 illustrates an example taken from the DISPLACE corpus, containing far-field speech signal and its annotations. The first panel contains the speaker-turn annotations, followed by its transcription in the second panel. In the first panel, s1, s2, and s3 represent the first, second, and third speakers of the conversations. The third panel shows the language labels.

### 3.1.4. Development and Evaluation set

For the challenge, the DISPLACE corpus was divided into development (Dev) and evaluation (Eval) sets. No training data was supplied, allowing participants to use any proprietary or publicly available resource to train their diarization systems.

The Dev set comprises 15.5 hrs (27 recordings) of bilingual conversations. Details of the Dev partition can be found in Table 5. Most conversations in the Dev set are around 30 minutes or 1 hr long. In the Dev set, a total of 46 male and 31 female speakers are present. The Eval set encompasses 16 hrs (29 recordings) of conversations in six different Indian languages, as well as Indian English. Most conversations in the Eval set have a duration of 30 minutes, except for three recordings that are 1 hr long. A total of 63 male and 26 female speakers are present in the Eval set. The set of speakers present in the Dev and Eval partitions are mutually exclusive. Additionally, the Eval conversations contain languages not seen in the Dev set.

### 3.1.5. DISPLACE Corpus Analysis

We present an analysis of the corpus by highlighting aspects of the speaker and language annotations.

Table 5: Details of the Development (Dev) set of the DISPLACE corpus.

| | # Speakers | | Code-swicth | Languages | Duration | #Recordings | | Total |
| | Male | Female | | | (hrs) | 30min | 60min | Duration (hrs) |
|---|---|---|---|---|---|---|---|---|
| | | | Bilingual | hi-en | 14.0 | 20 | 4 | |
| Dev Set | 46 | 31 | | ka-en | 1.0 | 2 | 0 | 15.5 |
| | | | | ma-en | 0.5 | 1 | 0 | |

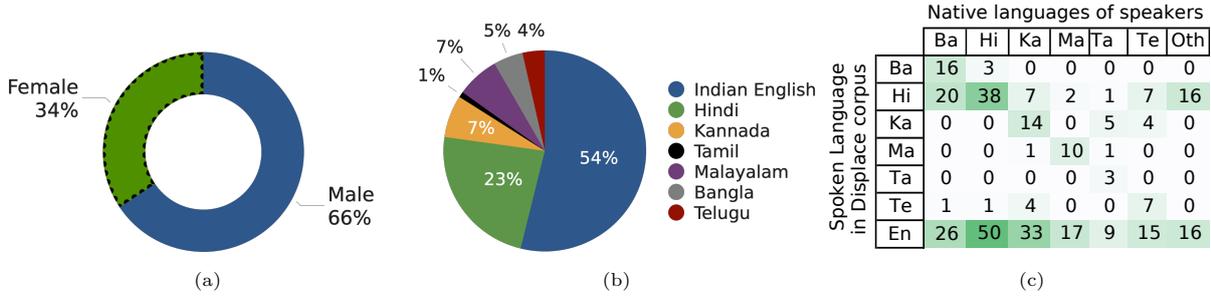

(a)　　　　　(b)　　　　　(c)

Figure 4: Distribution of (a) Speaker gender and (b) Speech duration per language in the DISPLACE dataset. (c) A count matrix illustrating the number of speakers with a native language (L1) for each of the spoken languages in the DISPLACE dataset.

1. *Speaker gender and spoken languages:*

   The DISPLACE dataset comprises a total of 166 unique speakers. The gender distribution is illustrated in Figure 4 (a). The DISPLACE corpus contains seven distinct languages in various combinations (as presented in Table 4). Figure 4 (b) illustrates the proportion (%) of speech duration for each of these spoken languages within the corpus. The dataset also demonstrates substantial diversity regarding the native languages of the speakers. Figure 4 (c) highlights the variation of the number of speakers for each of the spoken languages in the corpus with respect to the speakers' native language. A majority of the participants are proficient in at least one language other than their native language (L1) and English, which led to their involvement in conversations conducted in languages different from their L1. The insights derived from Figure 4 (b) and (c) affirm that the DISPLACE dataset covers a rich and diverse linguistic landscape.

2. *Speaker and language overlap:*

   The DISPLACE dataset was collected in an informal and spontaneous environment, resulting in a significant amount of speaker overlap where multiple speakers talk simultaneously. Figure 5 (a) illustrates the distribution of single speaker, speaker overlap, and non-speech segment durations. Approximately 9% of the entire dataset consists of non-speech regions, including silence and other non-speech acoustic categories, such as laughing, coughing, etc. Single speaker segments hold a major portion of the dataset (77%), followed by speaker overlap (with 14% share ≈ 5 hrs.). The speaker overlaps can involve speakers speaking in the same language (9%) or different languages (5%), as shown in the figure Figure 5 (a). Figure 5 (b) presents a distribution of the percentage of speaker overlap at the conversation level. This statistic is computed as a ratio of the total duration of speaker overlap within a conversation to the total speech duration of the recording. Similarly, Figure 5 (c) illustrates the percentage of language overlap at the conversation level. Here, language overlap represents the segments of speaker overlap with different languages.

   Based on these distributions, it can be concluded that the proportion of language overlap in the corpus is not substantial compared to speaker overlap.

3. *Dominant speakers and languages at the conversation level:*

   In the DISPLACE corpus, the conversations involve 3-5 speakers. Speakers in each conversation are ranked as first, second, third, and fourth dominant speakers based on their speech activity time. Figure 6 (a) illustrates the distribution of these dominant speakers in each conversation.

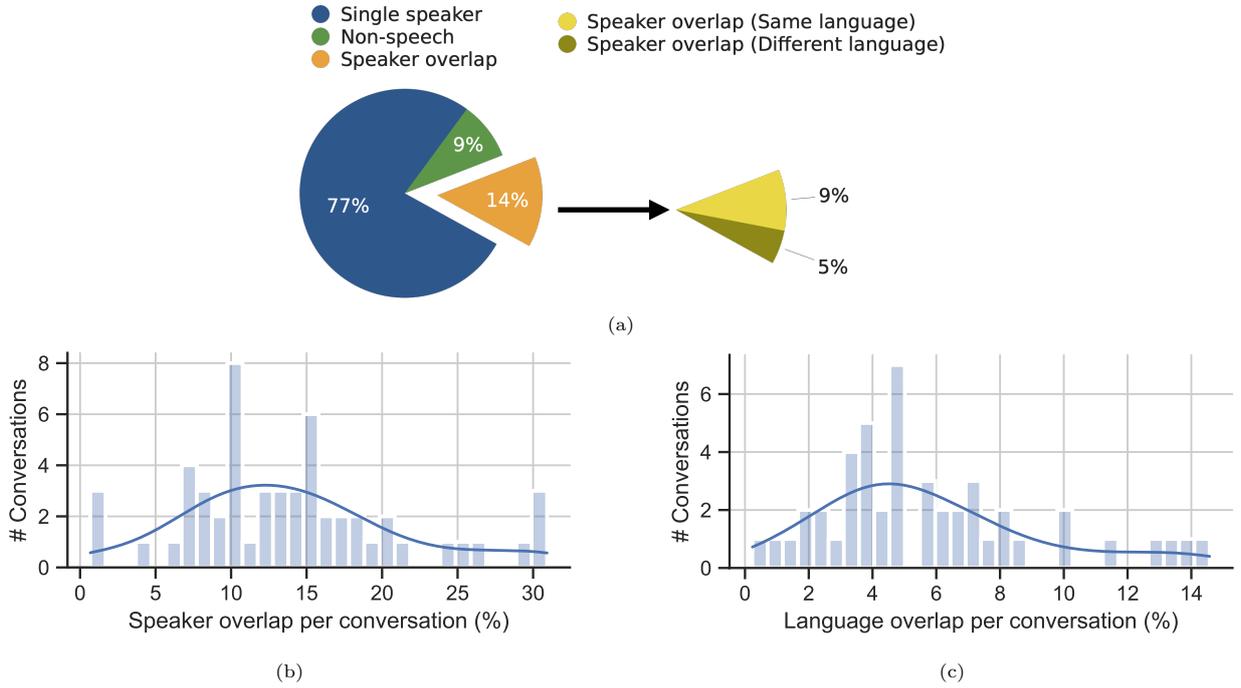

Figure 5: Speaker and language overlap statistics: (a) Pie chart illustrating non-speech, single speaker, and speaker/language overlap durations. A distribution of conversation-level (b) speaker overlap and (c) language overlap percentages for the DISPLACE corpus.

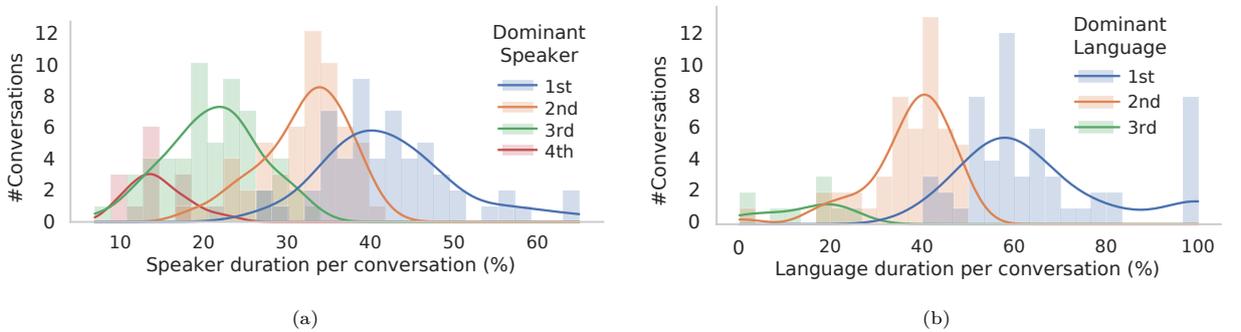

Figure 6: The percentage distribution of (a) the top 4 dominant speakers (in terms of spoken duration) and (b) top 3 dominant languages, at the recording level, over the entire dataset.

Figure 6 (b) shows the distributions of first, second, and third (where applicable) dominant languages per conversation. Language ranking is determined by the total duration of each language in a conversation. The eight monolingual recordings have a single language, represented by the bar at 100% in Figure 6 (b). For bilingual and trilingual conversations, the distributions of first and second dominant languages are closely spaced with some distribution overlap. This indicates that the DISPLACE is not biased towards any particular language within conversations. Trilingual conversations also have a relatively lower contribution from the third language.

4. *Speaker and language turns per conversation:*

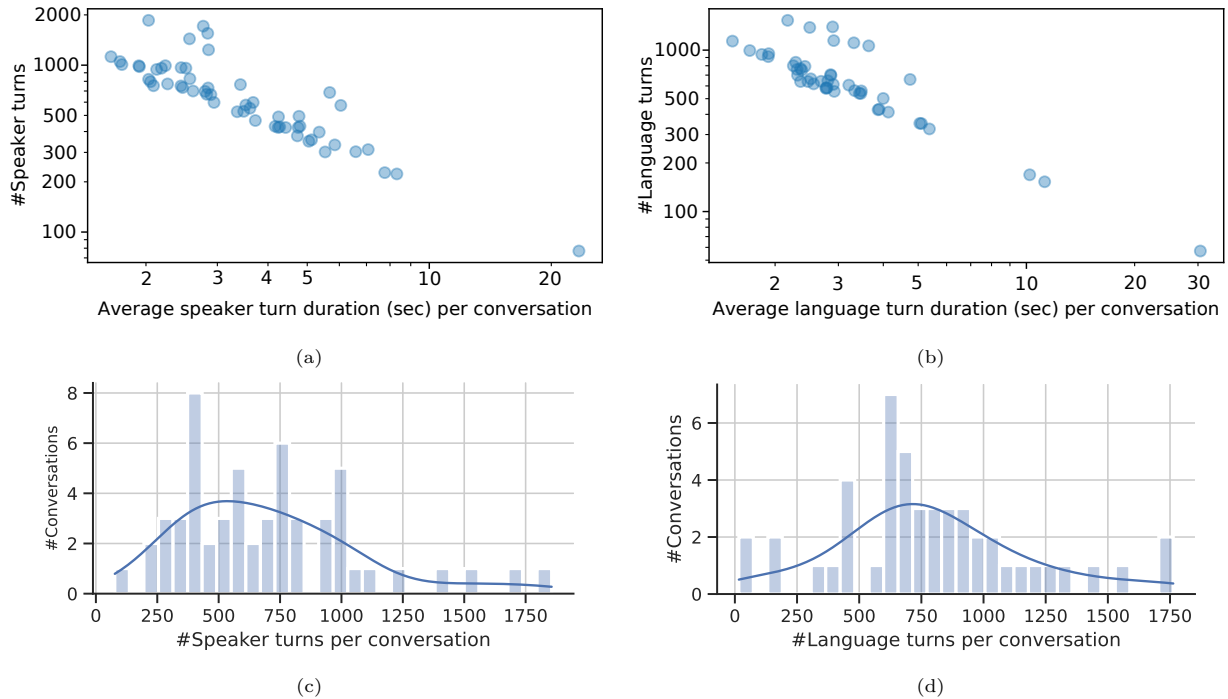

Figure 7: An illustration of the relationship between the number of speaker/language turns and turn durations. A scatter plot of conversation level (a) speaker segment duration versus number of speaker turns (b) language turn duration versus the number of language turns. The distribution of the number of (c) Speaker turns and (d) Language turns per conversations.

We define a *speaker turn* as a continuous speaking segment by a speaker that can include pauses ($\geq$ 500ms), provided the same speaker articulates before and after the pause. Similarly, we define a *language turn* as a contiguous segment during which a specific language is spoken. This duration can include pauses ($\geq$ 500ms) as long as the same language is spoken before and after the pause.

We exclude non-speech intervals when calculating both speaker and language turn durations. Figures 7 (a) and (b) depict the scatter between the number of speaker/language turns (on the y-axis) and the average durations of speaker/language turns at the conversation level (on the x-axis). In contrast, Figures 7 (c) and (d) present the distribution of the number of speaker and language turns (at the conversation level), respectively, across the DISPLACE corpus.

These statistics show that the average duration of speaker turns at the recording level can vary widely, ranging from 1.5s to 20s. Similarly, recording-level average language turn durations can range from 1.2s to 50s. The plots indicate that many language turns are considerably shorter than the recording-level averages and often consist of single, short words of duration 200ms.

*3.2. Challenge Tasks*

The DISPLACE challenge involves diarizing each conversational recording (far-field) by detecting speakers and language information automatically. In the challenge, participants were not provided with speech activity detection (SAD) systems, and hence, they were encouraged to develop their own SAD systems. The challenge contained two tracks:

- *Track-1: Speaker Diarization in multilingual scenarios*
  This track was designed to perform speaker diarization (identifying who spoke when) in multilingual conversational audio data.

- *Track-2: Language Diarization in multi-speaker settings*
  The objective of Track-2 was to conduct language diarization (identifying which language was spoken when) in multi-speaker conversational audio data.

Evaluations for both tracks considered only speech activity regions, including voiced back channels and fillers. Non-speech speaker activities, such as laughing, clapping, sneezing, etc., were excluded as non-speech during the evaluation. Moreover, small pauses[10] (with a duration of $\leq 300$ ms) made by a speaker were not considered as segmentation breaks, and were included as a part of the continuous segment.

### 3.3. Evaluation Phases and Metric

The evaluation was conducted in two phases: Phase-1 and Phase-2. In Phase-1, the evaluation was performed on a subset of the complete Eval set, comprising 20 recordings spanning 11.5 hrs. On the other hand, the complete Eval set was used for Phase-2 evaluation. The participation in Phase-2 evaluation was optional.

The system evaluation was performed using single-channel far-field conversational audio. The systems submitted to both tracks were compared with human reference segmentation. For the challenge, a Rich Transcription Time Marked (RTTM) file containing reference segmentation was generated for each conversation by combining close-talking annotations. Separate RTTM files for the SD and LD tasks were released to all the participating teams for the development dataset. The evaluation was conducted using the Diarization Error Rate (DER) [31] as the overall performance metric. For the challenge, DER was computed with overlap and without a collar.

## 4. Baseline Systems

A brief description of the sub-modules utilized in the baseline systems is provided below.

### 4.1. Speech Activity Detection (SAD)

The SAD system employed in the DIHARD-III baseline [31] is adopted in the DISPLACE challenge. The SAD model uses a Time Delay Neural Network (TDNN) architecture based on the Kaldi recipe. This model includes Time Delay Neural Network (TDNN) layers [49], followed by statistics pooling layers [50]. The model was trained using 40-dimensional Mel-frequency cepstral coefficients (MFCC) features extracted from the DIHARD-III development set to distinguish between speech and non-speech segments. MFCC features were extracted for a 25 ms frame with a shift of 10 ms. The training process was carried out for 40 epochs. More details are available in the DIHARD-III description [31].

### 4.2. Speaker Diarization (SD)

The initial step involves splitting the SAD speech regions into short overlapping segments of 1.5s with a 0.25s shift. These segments are then utilized to extract speaker embeddings (x-vectors) followed by probabilistic linear discriminant analysis (PLDA) scoring and clustering. We experimented with two different clustering approaches: Agglomerative hierarchical clustering (AHC) and Spectral clustering (SC). Finally, the clustering output is subsequently enhanced using the variational Bayes hidden Markov model (VB-HMM) with posterior scaling [51, 52].

The x-vectors are extracted from a 13-layer Extended-Time Delay Neural Network (ETDNN) [53]. The ETDNN model takes 40-D mel-spectrogram features as input, and generates 512-D x-vectors. The mel-spectrograms were extracted with a window of 25ms frames with a 10ms shift. The model was trained on the VoxCeleb1 [54] and VoxCeleb2 [55] datasets for the speaker identification task, aiming to distinguish among $7,146$ speakers. The x-vectors were processed through centering and whitening using statistics derived from the DISPLACE Dev set, followed by unit-length normalization. The transformed x-vectors obtained from a subset of VoxCeleb1 and VoxCeleb2 were used to train the PLDA model. The AHC approach is applied to the PLDA scores for obtaining speaker clusters. On the other hand, the SC algorithm takes processed PLDA scores as input, which are obtained by employing a sigmoid with a temperature scaling factor of 0.1.

---

[10]A pause is described as a time duration during which a speaker refrains from any type of vocalization. Vocalization encompasses speech, vocal sounds such as laughter, breathing, coughing, sneezing, lip-smacks, non-lexical sounds (e.g., "ahh," "umm," "uh-umm," "hmm," "huh," "ohh," etc.), or any other sound produced using the human sound production system.

Table 6: Baseline speaker diarization results (in terms of DER %) for the DISPLACE development (Dev) and evaluation (Eval) data (Phase-1 and Phase-2) using AHC and spectral clustering (SC), followed by VB-HMM re-segmentation.

|    | Method     | Dev   | Eval    |         |
|----|------------|-------|---------|---------|
|    |            |       | Phase-1 | Phase-2 |
| B1 | AHC        | 30.43 | 40.58   | 38.86   |
|    | AHC+VB-HMM | 28.91 | 39.78   | 37.96   |
| B2 | SC         | 27.87 | 33.16   | 32.51   |
|    | SC+VB-HMM  | **27.33** | **32.51** | **32.18** |

To improve the accuracy of speaker boundaries, the VB-HMM is initialized individually for each audio file using the clustering output. The hyper-parameters for the clustering algorithms and VB-HMM were fine-tuned to minimize the DER on the DEV set. We have two baseline systems for the SD track: AHC+VB-HMM (B1) and SC+VB-HMM (B2). The DER results for both B1 and B2 are presented in Table 6. It can be observed that baseline system B2 outperforms B1 on all three sets in terms of DER.

Table 7: The B2 (SC+VB-HMM) baseline speaker diarization results (in terms of individual components of DER %) for the DISPLACE Dev and Eval data (Phase-1 and Phase-2). The performance is also reported in terms of DER* (%) and DER** (%). Here, DER* represents the diarization error computed without a collar and excluding speaker overlap regions, while DER** denotes the DER when applying a 0.25s collar and excluding regions with speaker overlap.

|  | Dev/Eval | DER (%) | | | | DER* (%) | DER** (%) |
|--|----------|------|------|-----------|-------|----------|-----------|
|  |          | Miss | FA   | Confusion | Total | Total    | Total     |
|  | Dev      | **20.30** | 3.43 | 3.60 | 27.33 | 16.07 | 10.47 |
| Eval | Phase-1 | **22.70** | 2.71 | 7.10 | 32.51 | 20.52 | 15.10 |
|      | Phase-2 | **22.50** | 3.10 | 6.58 | 32.18 | 20.96 | 14.58 |

To pinpoint the primary source of errors in the DER, we examine the individual components of the DER metric as shown in Table 7, specifically for the best-performing baseline system, B2. It is evident that the miss component of the DER is notably higher when compared to the FA and confusion errors. The miss encompasses errors due to both SAD and overlap. To understand the main contributing factor to the high miss error, we evaluated diarization error without overlapped speech regions, denoted by DER* in table 7. For all the sets of DISPLACE data, the DER* values are lower than the corresponding DER errors by 12%. Thus, 12% error in the miss component is because of the overlap, and the remaining 10% error is due to SAD. To gain insight into how the systems perform under less stringent conditions, we also assess the DER metric using a collar of 0.25s and by disregarding regions of speech overlap, which we denote as DER**. Further, a 5% reduction is observed in the diarization error compared to their corresponding DER* values. This trend is consistent for all the splits of the DISPLACE corpus.

*4.3. Language Diarization (LD)*

The LD baseline system involves segment-level language embedding extraction, followed by clustering. The language embeddings are derived from the ECAPA-TDNN [56] based SpeechBrain [57] language recognition model[11]. The ECAPA-TDNN model is trained on the Voxlingua107 [58] corpus, comprised of 6628 hrs of speech data spoken in 107 different languages. The Voxlingua107 corpus contains short speech segments extracted from YouTube videos.

---
[11] https://huggingface.co/speechbrain/lang-id-voxlingua107-ecapa

Table 8: Baseline language diarization results (in terms of DER %) for the DISPLACE development (Dev) and evaluation (Eval) data (Phase-1 and Phase-2) using AHC and spectral clustering.

| Method | Dev | Eval | |
|---|---|---|---|
| | | Phase-1 | Phase-2 |
| B1 (AHC) | 47.22 | 41.82 | 41.92 |
| B2 (SC) | **46.95** | **41.22** | **41.67** |

The speech segments obtained from the SAD model are subsequently split into short overlapping segments of 0.4s with a 0.2s shift. The 256-D language embeddings are extracted for each segment, followed by pairwise cosine similarity score matrix computation and clustering. The LD baseline system also investigates two different clustering approaches: AHC and SC. The performance of the LD baseline system in terms of DER is reported in table 8. Nearly equivalent performance is noted for both clustering methods, though with a slight improvement observed for the SC approach. This pattern remains consistent across all splits of the DISPLACE corpus.

Table 9: The B2 (SC) baseline language diarization results (in terms of individual components of DER %) for the DISPLACE Dev and Eval data (Phase-2). The performance is also reported in terms of DER* (%) and DER** (%). Here, DER* represents the diarization error computed without a collar and excluding language overlap regions, while DER** indicates the evaluation result when applying a 0.25s collar and excluding regions with language overlap.

| Dev/Eval | | DER (%) | | | DER* (%) | DER** (%) |
|---|---|---|---|---|---|---|
| | | Miss | FA | Confusion | Total | Total | Total |
| | Dev | 12.50 | 3.45 | **31.00** | 46.95 | 46.30 | 41.76 |
| Eval | Phase-1 | 12.46 | 2.66 | **26.10** | 41.22 | 39.69 | 34.34 |
| | Phase-2 | 12.27 | 3.00 | **26.40** | 41.67 | 40.27 | 34.72 |

To further dissect the primary factor responsible for the overall DER, the DER metric is broken down into its constituent elements: miss, false alarms (FA), and confusion. Table 9 displays the components of DER obtained using the B2 (SC) method for both the Dev and Eval (Phase-1 and Phase-2) splits. The notably high confusion errors observed in both dataset splits strongly suggest that this is the main contributing element to the overall DER. The LD baseline system is also evaluated in much more relaxed settings in terms of DER* and DER**. A minimal decrement is observed in the DER* when compared to the corresponding DER results. This indicates that the influence of language overlap on the overall DER is minimal for the DISPLACE dataset. This is because of the small amount of language overlap present in the dataset. On the other hand, the DER** value achieved in the Phase-2 split is approximately 6-7% lower than the corresponding DER error. This pattern holds true across all splits. Having a collar of 250ms excludes the short duration ($\leq$ 500ms) foreign words from the evaluation. In the DISPLACE corpus, the average duration of the language segment is around 500ms (refer to Figure 2 (a) of [23]). These findings imply that the primary error arises from the short duration of the code-mixed instances.

## 5. Track-1: Submitted systems overview

A total of 13 teams (excluding the baseline system) submitted their systems in the Track-1 leaderboard. Out of which, 6 teams submitted their system reports providing details of their approach. This section presents a summary of the submitted systems. We refer the teams by their Team IDs based on their positioning on the Eval (Phase-1) leaderboard[12].

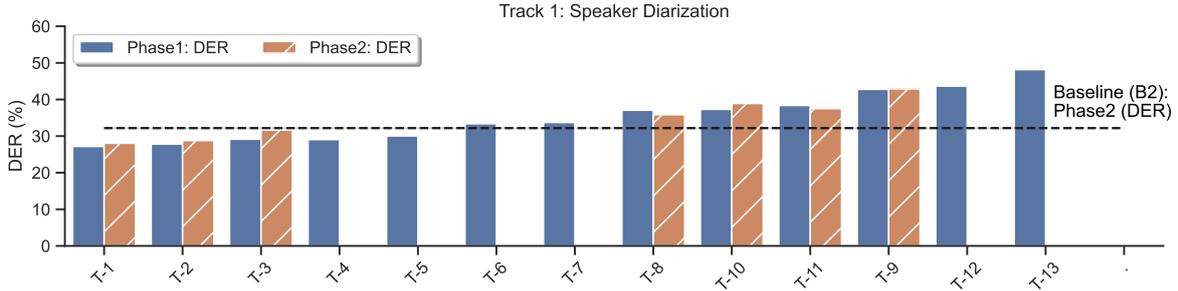

Figure 8: Performance evaluation of the submitted systems for the speaker diarization track. The performance is reported in terms of DER (%). Note that the participation in Phase-2 evaluation was optional. Hereby, only Phase-1 results are reported for T-4, T-5, T-6, T-7, T-12 and T-13.

- **SAD model:** The teams explored different SAD models for this track. T-1 utilized the pre-trained Silero VAD v4[13] model. T-2 explored three SAD models: Baseline SAD (BL-SAD), Silero VAD, and their in-house VAD. The final submission of T-2 includes a fusion of four systems utilizing BL-SAD and Silero VAD for speech activity detection. T-3 system used SAD model[14] from pyannote [59]. T-4 explores the Multilingual SAD module used in the SD system of the NeMo [60] toolkit. The model is based on MarbleNet [61]. T-10 employed a frame energy-based VAD, which compares the noise floor and frame energy to detect speech frames. However, An End-to-End Neural Diarization (EEND) [62] based approach, which encompasses SAD as a sub-task, is also investigated for this track by T-6 and T-4.

- **Speaker Embeddings:** The participating teams explored a wide range of open-source embedding extractors. A majority of teams used X-vector [53, 63] (T-2, T-3, T-10) and ECAPA-TDNN [56, 57, 64] (T-2, T-4, T-6) based embeddings. Other explored models are TitaNet [65] (T-4, T-6), pyannote [59] (T-6), RawNet3 [66] (T-6), Resnet-34 [67] and Resnet-293 (T-1). The T-6 compared pyannote, TitaNet, RawNet3, and ECAPA-TDNN extractors and observed ECAPA-TDNN as the best performer.

- **Clustering:** All the teams employed either AHC (T-1, T-3, T-4, T-6, T-10) or SC [68] (T-1, T-2, T-3, T-10) approaches to obtain speaker clusters. Some teams also performed re-segmentation using VB-HMM (T-2) and VBx (T-3).

*5.1. Performance*

A total of 5 out of the 13 teams demonstrated superior performance compared to the B2 baseline SD system. The performance of all 13 teams is summarized in figure 8. The team IDs are sorted based on their performance in the Phase-1 evaluation. The horizontal black (dashed) line represents the DER for the B2 baseline system on Eval (Phase-2). The first and second bars for each team represent DER values for the Phase-1 and Phase-2 evaluations. Note that participation in the Phase-2 evaluation was optional, resulting in some teams (e.g., T-4 [69], T-5, T-6, T-7, T-12, and T-13) not submitting their system results for Phase-2. T-1 achieved the lowest DER of 28.04, followed by T-2 (DER: 28.79) and T-3 (DER: 31.64) on the Eval Phase-2 set.

## 6. Track1: Top performers

A summary of the top-performing teams for track-1 is presented in table 10.

---
[12]https://codalab.lisn.upsaclay.fr/competitions/10588
[13]https://github.com/snakers4/silero-vad
[14]https://github.com/desh2608/diarizer/blob/1bfcd0771fdcca81f318f230ab208116de8d9dda/diarizer/vad/pyannote_vad.py#L5

*6.1. Team-1 (T-1):*

The team investigated a pre-trained Silero-v4 VAD, Resnet architecture-based embedding extractors (Resnet-34 and Resnet-293), and SC-based clustering. The embedding extractors were trained on VoxCeleb-2 and Common Voice 12.0 corpora. An 80-D Mel filter-bank features, extracted from a 25ms window with a 10ms shift, were fed as input to the model. The system used the MUSAN dataset [70] for data augmentation by adding music, noise, and reverberation from the RIR dataset. The models were trained using AAM-Softmax Loss with 150 epochs. The approach experimented with two segment sizes (1.5s and 2s) and three shifts (0.4s, 0.5s and 0.75s). The lowest DER on the Dev set was obtained for 2s segment size with 0.4s shift. The team compared SC and AHC approaches and observed SC to be the outperforming method. The final submission includes Silero-v4 VAD with a 0.15 threshold, Resnet-293 extractor with 2s segments, and 0.4s shift and SC, along with overlap handling. Based on the experimental results, the team observed that the VAD model needs to be improved to decrease the final DER.

Table 10: Track 1: Summary of the top three performing submitted systems for SD task.

| Team ID | Implementation Details | | | | | | DER (Eval) | |
|---|---|---|---|---|---|---|---|---|
| | Training data | SAD | Embedding Extractor | Clustering [Re-seg.] | Ovrlp handle | Segment size [shift] | Fusion | Phase1 | Phase2 |
| T-1 | VoxCeleb2-dev, Common Voice (12.0) | Silero-v4 | Resnet-293 | SC[-] | ✓ | 2s[0.4s] | - | 27.18 | 28.04 |
| T-2 | VoxCeleb 1, VoxCeleb 2 | BL-SAD, Silero | Xvector, ECAPA-TDNN | PLDA+SC[-], SC[VBHMM] | - | 1.5s[0.25s], 1.5s[0.75s] | ✓ | 27.81 | 28.79 |
| T-3 | AMI, AliMeeting, VoxCeleb1, CN-CELEB | pyannote | x-vectors (ResNet101) | SC[VBx] | - | 1.5s[0.75s] | ✓ | 29.14 | 31.64 |

*6.2. Team-2 (T-2)*

The team used different combinations of pre-trained SAD (BL-SAD and Silero), embedding extractors (X-vector and ECAPA-TDNN) and clustering (SC) methods. VoxCeleb-1 and VoxCeleb-2 datasets, augmented with noise and reverberation, were used for training embedding extractors. The MUSAN and RIR datasets were utilized to simulate noise and reverberation samples, respectively. For ECAPA-TDNN, mean-normalized (performed at segment level) 80-D log Mel-filterbank energies are passed as input to the model and obtain 192-D embeddings. The embeddings are extracted for 1.5s segments with two shifts: 0.25s and 0.75s. The top four performing systems of the team include S1 (BL-SAD+ECAPA-TDNN+SC+VB-HMM and S2 (BL-SAD + ECAPA-TDNN + SC + VB-HMM), S3 (Silero-SAD+ECAPA-TDNN+SC+VB-HMM) and S4 (BL-SAD+X-vector+PLDA+SC). The fusion of S1, S2, S3, and S4 is the final submission of the team.

*6.3. Team-3 (T-3)*

The T-3 system employed SAD model[15] from pyannote [59]. The speech recordings are split into 1.5s segments with 0.75s shift. The embeddings are extracted using a pre-trained SincNet-based extractor [63] containing a ResNet101 [67] architecture. The input to the model is 64-D log Mel-filter bank features, and the model generates 256-D x-vectors as output. The x-vectors are clustered using the SC method, followed by VBx [63] re-segmentation. The system also experimented with AHC-based clustering individually and in combination with VBx. The T-3 system used AMI, AliMeeting, VoxCeleb1, and CN-CELEB datasets for training and development purposes. The final submission includes pyannote SAD, x-vectors derived from ResNet101, SC, and VBx re-segmentation.

---

[15]https://github.com/desh2608/diarizer/blob/1bfcd0771fdcca81f318f230ab208116de8d9dda/diarizer/vad/pyannote_vad.py#L5

*6.4. Top three teams: Performance comparison*

This sub-section presents a performance comparison of the top three teams in terms of DER (individual components), DER* and DER**. When comparing the DER and DER* values across all teams, a notable ($\approx 10\%$) decrease in the diarization error is evident. Thus, it can be inferred that overlap miss error is the main contributor to the overall DER. These observations are aligned with the trend noted in the SD baseline (B2) systems (sub-section 4.2). When comparing the DER* and DER** values across all teams, a further decrement ($\approx 8\%$) in the diarization error is noted. These findings lead to the conclusion that imprecise speaker boundaries and speaker overlap are the two primary contributing factors to the overall DER.

Table 11: The speaker diarization results (in terms of DER (%), DER*(%) and DER** (%)) for the top three performing submitted systems. The results are reported for the DISPLACE Eval (Phase-2) set. Here, DER* represents the diarization error computed without a collar+ignoring speaker overlap, while DER** indicates the result with a 0.25s collar+ignoring speaker overlap.

| Team IDs | DER (%) | | | | DER* (%) | DER** (%) |
|---|---|---|---|---|---|---|
| | Miss | FA | Confusion | Total | Total | Total |
| T-1 | **17.24** | 6.30 | 4.50 | 28.04 | 16.33 | 7.27 |
| T-2 | **18.23** | 5.93 | 4.63 | 28.79 | 17.94 | 9.64 |
| T-3 | **18.90** | 6.62 | 6.12 | 31.64 | 21.04 | 12.39 |
| B2 | **22.50** | 3.10 | 6.58 | 32.18 | 20.96 | 14.58 |

# 7. Track-2: Submitted systems overview

A total of 6 teams (excluding the baseline system) participated in the Track-2 leaderboard. However, only three teams submitted their system reports outlining the specifics of their approach. This section provides a summary of the submitted systems.

- **Conventional Approach:** The T-5 system employed a conventional diarization approach, which involves a SAD model from pyannote [59], ECAPA-TDNN [56] based embedding extractor and AHC clustering along with the representation expansion strategy. This approach consists of calculating a cosine similarity matrix for pairs of items, selecting the top-K embeddings based on their similarity scores, and finally computing the average embedding from the selected set of top-K embeddings. The average embedding is considered as the representative of the current embedding. The motivation for using the representation expansion strategy is to enhance the resilience of the embeddings used for clustering.

- **End-to-End Models based Approach:** The language identification module of the Whisper [71] toolkit is used in the T-1 and T-2 systems. The output from the language model is post-processed to get a lower dimensional representation, which is used to get the final diarization output. Note that this approach is not based on clustering but rather a language detection driven LD method. Team T-3 also used an end-to-end approach with a wav2vec [72] based language embedding extractor.

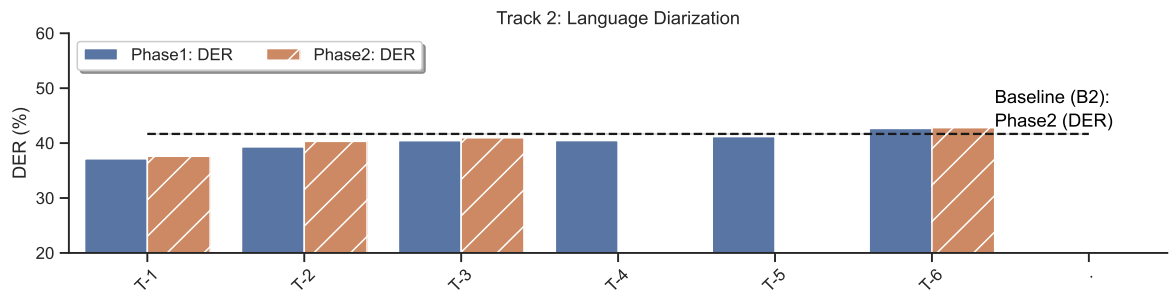

Figure 9: Performance evaluation of the submitted systems in the language diarization track. The performance is reported in terms of DER (%). Note that the participation in Phase-2 evaluation was optional. Hereby, only Phase-1 results are reported for T-4 and T-5.

*7.1. Performance*

Figure 9 provides a concise overview of the performance comparison among all participating teams in track-2[16]. The horizontal dashed (black color) line represents the B2 baseline DER value. Team T-1 stands out with the highest performance in Eval Phase-2 (DER: 37.60), followed by T-2 (DER: 40.32) and T-3 (DER: 40.97). For Phase-1, teams T-3 (DER: 40.44) and T-4 (DER: 40.46) exhibit nearly comparable performance. Teams T-4 and T-5 did not participate in the Phase-2 evaluation. Team T-6 achieved a DER of 42.81 in the Phase-2 evaluation. With the exception of T-6, all teams outperformed the LD baseline system.

## 8. Track2: Top performers

This section briefly describes the top three systems with their summary in table 12.

*8.1. Team-1 (T-1)*

The T-1 [73] proposed an end-to-end multi-resolution approach for the LD task using the Whisper decoder. An accent identification system was employed as an attention technique to decrease the number of output languages obtained from Whisper Language Identification module. The Whisper L1 model was tested on 11 different segment sizes starting from 1s to 11s with an increment of 1s, and obtained diarization output for each segment size. The lowest DER is obtained for 5s window size on the Dev set of the challenge. The team combined diarization outputs of all the segment sizes using the DOVER-Lap approach and witnessed a significant improvement in the DER.

*8.2. Team-2 (T-2)*

The T-2 system also utilized the Whisper model for the LD task. The pre-trained Whisper model is used to get the language labels, which are further divided into English and non-English categories. The English and non-English language labels are used to get the final diarization output.

*8.3. Team-3 (T-3)*

The T-3 system also employed an end-to-end E2E framework inspired by the approach proposed by Liu et al. [46]. The T-3 used the wav2vec model as an embedding extractor. The training of the wav2vec model involved two strategies: pre-training and fine-tuning. During the pre-training phase, contrastive predictive coding (CPC) was applied to unlabeled audio data for predicting features at the syllable/word level. This pre-training was carried out using data from 23 Indic languages. In the subsequent fine-tuning phase, the objective was to enable the model to distinguish between different languages. The Dev set was utilized for fine-tuning using connectionist temporal classification (CTC) loss. The 768-D embeddings from the wav2vec model were extracted from each 5s speech segment with a 4s segment shift.

---

[16]We re-evaluated the Phase-1 performance for teams that participated in Phase-2 evaluation, using the updated system output.

Table 12: Track 2: Summary of the top three performing submitted systems for LD task.

| Team ID | Implementation Details | | | | DER (Eval) | |
|---|---|---|---|---|---|---|
| | **Embedding Extractor** | **Multi-resolution approach** | **Segment size [shift]** | **Fusion strategy** | **Phase1** | **Phase2** |
| T-1 | Pre-trained Whisper | ✓ | 1s-11s[1s] | DOVER-Lap | 37.14 | 37.60 |
| T-2 | Pre-trained Whisper | ✗ | - | - | 39.31 | 40.32 |
| T-3 | fine-tuned Wav2Vec | ✗ | 5s[4s] | ✗ | 40.44 | 40.97 |

*8.4. Top performing teams: Performance comparison*

Table 13 presents a performance comparison among the top-ranking teams in terms of individual components of the DER, DER*, and DER** metrics. When we analyze the individual DER components for all three teams, it becomes evident that confusion errors are the primary contributing factors to the overall error. We observe a negligible decrease when comparing the total DER and DER*. However, all three teams exhibit DER** values which was 10% lower than the DER error. This indicates that allowing a tolerance of 0.25s across actual language boundaries significantly reduces diarization error. These patterns align with the observations made for the LD baseline system.

## 9. Open Challenges and Future Scope

The DISPLACE challenge, being first of its kind, also had certain limitations. This section provides insights into some of these shortcomings based on the observations made from both the baseline and submitted systems.

- *Overlap handling in Speaker Diarization:*

  The significant issue faced by the speaker diarization system is the presence of speech overlap, which is a major challenge. The baseline, T-2, and T-3 systems lack a dedicated mechanism to handle overlaps, resulting in a higher DER miss error. Despite T-1 handling overlapped regions in its own manner, its DER miss value is on par with the DER miss values of T-2 and T-3. This leads us to conclude that explicitly addressing overlapped speech is essential to effectively reduce the overall diarization error.

- *High DER confusion errors in Language Diarization:*

  The LD systems, including baseline and top-performing teams, encountered challenges in accurately identifying language boundaries and distinguishing between various languages. This difficulty is substantiated by the elevated confusion error rates in the DER. Several factors contribute to this issue. The presence of shorter duration code-mixed instances is the primary sources of diarization error.

Table 13: The language diarization results (in terms of DER** (%) and individual components of DER %) for the top three performing submitted systems. The results are reported for the DISPLACE Eval (Phase-2) set. Here, DER* represents the diarization error computed without a collar+ignoring language overlap, while DER** indicates the result with a 0.25s collar+ignoring language overlap.

| Team IDs | DER (%) | | | | DER* (%) | DER** (%) |
|---|---|---|---|---|---|---|
| | **Miss** | **FA** | **Confusion** | **Total** | **Total** | **Total** |
| T-1 | 4.70 | 8.3 | **24.40** | 37.42 | 35.73 | 26.57 |
| T-2 | 4.80 | 8.2 | **27.20** | 40.19 | 38.91 | 30.90 |
| T-3 | 4.87 | 8.2 | **27.90** | 40.97 | 39.64 | 31.48 |
| B2 | 12.27 | 3.00 | **26.40** | 41.67 | 40.27 | 34.72 |

- *ASR on the code-mixed dataset:*

  Speaker and language diarization are the two important pre-processing steps for ASR systems when dealing with code-mixed/switched datasets. In the future, this work can be extended by conducting ASR on the DISPLACE dataset, leveraging the results from speaker and language diarization.

- *Expansion of the DISPLACE corpus:*

  In the future, we intend to expand the DISPLACE corpus by including additional code-mixed/switched informal conversations involving multiple speakers.

## 10. Summary

This paper presents a comprehensive overview of the DISPLACE Challenge 2023, designed to promote research in the field of processing multilingual, multi-speaker conversational audio. The challenge had two tracks: one focusing on speaker diarization within multilingual contexts and the other on language diarization within multi-speaker settings, evaluated using the same dataset. Prior challenges and research in speaker diarization have predominantly concentrated on scenarios with a single language in the recordings. Similarly, previous language diarization studies have typically used recordings featuring a single speaker. To the best of our knowledge, no publicly available dataset matches the diverse characteristics observed in the DISPLACE dataset, which includes elements such as multilingual content, code-mixing, multi-speaker interactions, and natural conversational speech.

The paper outlines the unique features of the DISPLACE dataset, including its data collection, annotation procedures, and corpus analysis. It also outlines the baseline system and briefly discusses the systems submitted, covering aspects such as the SAD model, embedding extractor, and clustering methods. The paper conducts a performance comparison of all teams and baseline systems. The paper offers an in-depth examination of the top three performing systems in both tracks and investigates the primary sources of errors leading to DER values. Despite extensive global efforts in system development, the dataset remains challenging in both tracks, underscoring the need for continued research to advance speech technology within the realm of natural multilingual conversations.

**Acknowledgments**

The authors would like to thank the Ministry of Electronics and Information Technology (MeitY), Government of India for providing financial support through the National Language Translation Mission (NLTM): BHASHINI project, SP/MITO-22-001 grant. The authors extend their appreciation to Sidharth, Ranjana H, and Swapnil Padhi for their valuable assistance in the process of DISPLACE data collection and annotation enhancement.